 \documentclass[onecolumn,12pt,draft]{IEEEtran} 

\usepackage{epsf}

\begin{document}

\title{ From Sigmoid Power Control Algorithm to Hopfield-like Neural Networks: 
``SIR''-Balancing Sigmoid-Based Networks- Part II: Discrete Time
} 

\author{Zekeriya Uykan \thanks{Z. Uykan is with Helsinki University of Technology, Control 
Engineering Laboratory, FI-02015 HUT, Finland. E-mail: zekeriya.uykan@hut.fi.  
The author is a visiting scientist at Harvard University Broadband Comm Lab., Cambridge, MA,
and this work has been performed during his stay at Harvard University.} 
}


\maketitle
\begin{abstract}
In the first part in \cite{Uykan08a}, we present and analyse a Sigmoid-based 
"Signal-to-Interference Ratio, (SIR)" balancing dynamic 
network, called Sgm"SIR"NN, which exhibits similar properties as traditional Hopfield NN does, 
in continuous time. In this second part, we present the corresponding 
network in discrete time:  
We show that 
in the proposed discrete-time network, called D-Sgm"SIR"NN, the defined error vector approaches 
to zero in a finite step in both synchronous and asynchronous work modes.  
Our investigations show that 
i) Establishing an analogy to the distributed (sigmoid) power control algorithm in 
\cite{UykanPhD01} and \cite{Uykan04}
if the defined fictitious "SIR" is equal to 1 at the converged eqiulibrium point, then it is one of the 
prototype vectors. 
ii) The D-Sgm"SIR"NN exhibits similar features as discrete-time Hopfield NN does. 
iii) Establishing an analogy to the traditional 1-bit fixed-step power control algorithm, the corresponding 
"1-bit" network, called Sign"SIR"NN network, is also presented. 

\end{abstract}


\section{Introduction \label{Section:INTRO}}

\PARstart{T}{h}is paper is a continuation of the study in \cite{Uykan08a} where a continuous-time 
"Signal-to-Interference Ratio, (SIR)"-balancing neural network 
is presented which includes Hofield Network and sigmoid-based 
power control algorithm algorithm of \cite{UykanPhD01} and \cite{Uykan04} as special cases, both of whose  scopes of 
interest, motivations and settings are completely different. In this paper, we 
examine the discrete-time counterpart of \cite{Uykan08a}, and propose two discrete-time 
sigmoid-basis SIR-balancing networks which exhibit similar features which generaly are 
attributed to recurrent neural networks like discrete-time Hopfield Networks.

Hopfield Neural Networks 
has been an important focus of 
research area since early 1980s whose applications vary from combinatorial 
optimization (e.g. \cite{Matsuda98}, \cite{Smith98} among many others) 
including traveling salesman problem (e.g. \cite{Tan05}, \cite{Huajin04} among others) 
to image restoration (e.g. \cite{Paik92}), 
from various control engineering optimization problems including in robotics 
(e.g. \cite{Lendaris99} among others) to associative memory systems 
(e.g. \cite{Farrel90} among others), etc. 
For a tutorial and further references about Hopfield NN, see e.g. \cite{Zurada92} and 
\cite{Haykin99}.

\smallskip 

In the first part in \cite{Uykan08a}, we present a Sigmoid-based "Signal-to-Interference Ratio (SIR)" 
balancing dynamic 
network, called Sgm"SIR"NN, which exhibits similar properties as traditional Hopfield NN does,  
is presented and analysed in continuous time. In this second part, we present the 
corresponding networks in discrete time. 
Our investigations show that 
i) Establishing an analogy to the distributed (sigmoid) power control algorithm in \cite{UykanPhD01} 
and \cite{Uykan04}, 
if the defined fictitious "SIR" is equal to 1 at the converged eqiulibrium point, then it is one of the 
prototype vectors. 
ii) The D-Sgm"SIR"NN exhibits similar features as discrete-time Hopfield NN does. 
iii) Establishing an analogy to the traditional 1-bit fixed-step power control algorithm, the corresponding 
"1-bit" network, called Sign"SIR"NN network, is also presented.

The paper is organized as follows:  The proposed D-Sgm"SIR"NN and its 1-bit version network 
is presented and their stability features are analysed in 
section \ref{Section:D-SgmCIRNN}.  Simulation results are presented in Section 
\ref{Section:SimuResults}  
followed by Concluding Remarks in Section \ref{Section:CONCLUSIONS}. 

\smallskip

\section{``SIR''-Balancing Sigmoid-Based Networks in Discrete Time  \label{Section:D-SgmCIRNN}}

We start with the standard definition of Signal-to-Interferende+Noise-Ratio (SIR) in a 
cellular radio system, in which $N$ mobiles share the same channel (e.g. \cite{Zander92a}, \cite{Zander92b}).  

\begin{equation} \label{eq:cir}
\gamma_i = \frac{ g_{ii} p_i}{ \nu_i + \sum_{j = 1, j \neq i}^{N} g_{ij} p_j },  \quad i=1, \dots, N
\end{equation}

where  $p_i$ is the transmission power of mobile $i$, $g_{ij}$ is
the link gain from mobile $j$ to base $i$ involving path loss,
shadowing, multi-path fading (as well as the spreading/processing gain in case of 
CDMA transmission \cite{Rappaport96}, etc), and $\nu_i$ is the
receiver noise at base station $i$.  

Because, in power control, the positive transmit power can not be arbitarily 
small and large in practice, 
we write the eq.(\ref{eq:cir}) with the minimum and maximum power constraints as follows: 

\begin{equation} \label{eq:cirWithMaxTxp}
\bar{\gamma}_i = \frac{ g_{ii} \max \{ p_{min}, \min \{p_{max}, p_i \} \} }
 	{ \nu_i + \sum_{j = 1, j \neq i}^{N} g_{ij} \max \{ p_{min}, \min \{p_{max}, p_j \} \} },  
	\quad i=1, \dots, N
\end{equation}

where $p_{min}$ and $p_{max}$ is the minimum and maximum transmit powers. The SIR model in 
(\ref{eq:cirWithMaxTxp}) can be further written in a more generalized equation as follows 
using neural networks termonilogy 

\begin{equation} \label{eq:cirFx}
\bar{\gamma}_i = \frac{ g_{ii} y(p_i)}{ \nu_i + \sum_{j = 1, j \neq i}^{N} g_{ij} y(p_j) },  
		\quad i=1, \dots, N
\end{equation}

where $y(\cdot)$ represents the modeling of lower and upper bounding the transmit power and of any 
other effects e.g. power amplifier, etc. For example, 
$y(p_i) =  \max \{ p_{min}, \min \{p_{max}, p_i \} \}$ or corresponding piecewise linear function 
$y(p_i) = | p_i + p_{max} | - | p_i - p_{max} | $  yields eq.(\ref{eq:cirWithMaxTxp}). 

By relaxing the positivity conditions in the power control problem in (\ref{eq:cirFx}) and using sigmoid  
as the bounding function to the states in the denominator, and a different function in the nominator,  
the following fictitious "SIR" is defined in \cite{Uykan08a}:

\begin{equation} \label{eq:cirC}
\bar{\theta}_i = 
      \frac{ a_{ii} f_3(x_i)}{ b_i + \sum_{j = 1, j \neq i}^{N} w_{ij} f_2(x_j) },  \quad i=1, \dots, N
\end{equation}

where  
$\theta_i$ is the defined fictitious ``SIR'', 
$x_i$ is the state of the $i$'th neuron, 
$a_{ii}$ is the feedback coefficient from its state to its input layer,  
$w_{ij}$ is the weight from the output of the $j$'th neuron to the input of the 
$j$'th neuron, and $f_2(\cdot)$ represents the sigmoid function, 
and $f_3(\cdot)$ represents the function used for self-state-feedback. 
Sigmoid function is defined as 
$f_2(e_i) = 1 - \frac{1}{1 + exp(-\sigma_1 e_1)}$, 
where $\sigma_1>0$ is called slope of $f_2(\cdot)$, which is equal to 
its derivative with respect to its argument at the origin 0.

It's shown in \cite{Uykan08a} that choosing $f_3(\cdot)$ as a unity function in 
(\ref{eq:cirC}), i.e., $f_2(x_i) = x_i$, yields a network, 
called Sgm''SIR''NN which exhibits similar features as Hopfield NN does.
So, following fictitious "SIR" is defined 

\begin{equation}\label{eq:cirD}
\frac{\bar{\theta}_i}{\theta_{i}^{tgt}} = 
      \frac{ a_{ii} x_i}{ b_i + \sum_{j = 1, j \neq i}^{N} w_{ij} f_2(x_j) },  \quad i=1, \dots, N
\end{equation}

which is shown to satisfy the equilibrium points (prototype vectors) of the following 
dynamic network with ${\theta_{i}^{tgt}}=1$, called Sgm"SIR"NN in \cite{Uykan08a}:

\begin{equation}\label{eq:contSgmNN_v2}
\dot{ {\mathbf x}} =  {\mathbf f}_1 \Big( -{\mathbf A}{\mathbf x} + {\mathbf W} \mathbf{f_2(x)} 
	+ {\mathbf b} \Big)
\end{equation}

where $\dot{ {\mathbf x}}$ represents the derivative of ${\mathbf x}$ with respect to time and

\begin{equation} \label{eq:matA_W_b}
{\mathbf A} = 
\left[
\begin{array}{c c c c}
a_{11}   &   0   & \ldots  &  0 \\
0     &   a_{22} & \ldots  &  0 \\
\vdots &      & \ddots  &  0 \\
0     &   0   & \ldots  &  a_{NN}
\end{array}
\right], 
\quad \quad 
{\mathbf W} = 
\left[
\begin{array}{c c c c}
0  &   w_{12}   & \ldots  &  w_{1N} \\
w_{21}     &   0 & \ldots  &  w_{2N} \\
\vdots &     & \ddots  &  \vdots \\
w_{N1}    &   w_{N2}   & \ldots  &  0
\end{array}
\right]
\quad \quad
{\mathbf b} = 
\left[
\begin{array}{c}
b_1 \\
b_2 \\
\vdots \\
b_N
\end{array}
\right]
\end{equation}

In eq.(\ref{eq:matA_W_b}), ${\mathbf A}$ shows the self-state-feedback matrix with $a_{jj}>0$, ${\mathbf W}$ with zero 
diagonal shows the connection weight matrix from outputs to other neuron's inputs, and ${\mathbf b}$ is 
a threshold vector. 

It's shown in \cite{Uykan08a} that the network in (\ref{eq:contSgmNN_v2}) exhibits similar features as 
continuous Hopfield Network does. In this paper, we examine its discrete-time version.

From the fictitious CIR definition in eq.(\ref{eq:cirD}), 
let's define the following error signal

\begin{equation} \label{eq:DiffSgmNN_e_signal_elementwise_v2}
e_i = -a_{ii} x_i + I_i, \quad \quad 
\textrm{where} \quad I_i =  b_i \sum_{j = 1, j \neq i}^{N} w_{ij} f_2(x_j), 
	\quad i=1, \dots, N
\end{equation}

Writing (\ref{eq:DiffSgmNN_e_signal_elementwise_v2}) in matrix form gives 

\begin{equation}\label{eq:DiffSgmNN_e_signal_v2}
{\bf e} = -{\bf A}{\bf x} + 
                {\bf W}{\bf f}_2({\bf x}) + {\mathbf b}
\end{equation}

which is equal to the argument of the ${\mathbf f}_1( \cdot )$ in the network Sgm"SIR"NN 
in eq.(\ref{eq:contSgmNN_v2}).

From eq. (\ref{eq:contSgmNN_v2}) and (\ref{eq:DiffSgmNN_e_signal_v2}), $\dot{ {\bf x}} = {\bf e}$.  
If $e_i = 0$ given that $x_i \neq 0$ and $I_i \neq 0$, then, 
from eq.(\ref{eq:cirD}) and (\ref{eq:DiffSgmNN_e_signal_elementwise_v2}), 
$\hat{\theta}_i = \theta_i^{tgt} = 1$.  

The prototype vectors are defined as those ${\mathbf x}$'s 
which make $\theta_i = \theta_{i}^{tgt} = 1, \quad i=1, \dots, N$ given 
that $x_i \neq 0$ and $I_i \neq 0$. So, from (\ref{eq:cirC}) and (\ref{eq:cirD}), the prototype vectors 
make the error signal zero, i.e., $e_i=0, \quad i=1, \dots, N$. 

\subsection{Discrete Sgm"SIR"NN Network  \label{FDPC}}

In this section, we present a Sigmoid based "SIR"-balancing network which exhibits similar features as 
discrete Hopfield NN does.

Discretizing the differential equation (\ref{eq:contSgmNN_v2}) by the Euler method gives

\begin{equation} \label{eq:Diff_DPCA_Discrete}
{\mathbf x}^{k+1} =  {\mathbf x}^{k} - \alpha 
{\mathbf f}_1 \Big( -{\mathbf A}{\mathbf x}^{k} + {\mathbf W} \bf{f}( \bf{x}^{k} ) + {\mathbf b} \Big)
\end{equation}

where ${\mathbf A}, {\mathbf W}$ and ${\mathbf b}$ are defined as in eq.(\ref{eq:matA_W_b}), and 
$k$ represents the iteration step.

From eq.(\ref{eq:Diff_DPCA_Discrete}) and (\ref{eq:matA_W_b}), 

\begin{equation} \label{eq:Diff_SgmCIR_elementwise}
x_{j}^{k+1} = x_{j}^k + \alpha^k f_1 \Big( 
-a_{jj} x_j^k + b_j + \sum_{i = 1, i \neq j}^{N} w_{ij} f_2(x_i^k) \Big)  
\quad \quad \quad j=1, \dots, N
\end{equation}

where $\alpha^k$ is the step size at time $k$.

We will call the network in eq.(\ref{eq:Diff_SgmCIR_elementwise}) as D-Sgm"SIR"NN (Discrete Sigmoid 
``SIR''-balancing neural network). 

The performance index is defined as $l_1$-norm of the error vector in 
(\ref{eq:DiffSgmNN_e_signal_v2}) as follows

\begin{eqnarray} \label{eq:perfIndexLk} 
V(k) = || {\bf e}(k) ||_1  &  =  &  \sum_i^N | e_i(k) | \\
              		   &  =  &  \sum_i^N | -a_{ii} x_i + I_i |  
				\quad \textrm{where} \quad I_i = b_i + \sum_{j = 1, j \neq i}^{N} w_{ij} f_2(x_j) 
\end{eqnarray}

In what follows, we examine the evolution of the the energy function in (\ref{eq:perfIndexLk})
in synchronous and asynchronous work modes. 
Synchronous mode means that at every iteration step, at most only one state is updated, whereas 
asynchronous mode refers to the fact that all the states are updated at every iteration step 
according to eq.(\ref{eq:Diff_SgmCIR_elementwise}).

\vspace{0.2cm}
\emph{Proposition 1: } 
\vspace{0.2cm}

In asynchronous mode, 
in the D-Sgm"SIR"NN in eq.(\ref{eq:Diff_SgmCIR_elementwise}) with a symmetric matrix ${\bf W}$, 
the $l_1$-norm of the error vector in eq.(\ref{eq:perfIndexLk}) decreases at every step 
for a nonzero error vector, 
i.e., the error vector goes to zero for any $\alpha^k$ 
such that 

\begin{equation} \label{eq:lemma1_if_alpha}
|e_j^k| > |a_{jj} \alpha^k f_1(e_j^k)|
\end{equation}

if 

\begin{equation} \label{eq:lemma1_if}
| a_{jj} | \geq k_2 \sum_{i=1, (i \neq j)}^{N}  |w_{ij}|
\end{equation}

where $k_2= 0.5 \sigma$ is the the global Lipschitz constant
of $f_2(\cdot)$  as shown the in Appendix A.


\vspace{0.2cm}


\begin{proof} \label{prf:1}

\vspace{0.2cm}

In asynchronous mode, only one state is updated at an iteration time. 
Let $j$ shows the state which is updated at time $k$ whose error signal is different 
than zero, i.e., $e_j = -a_{jj} x_j + I_j \neq 0, \quad \quad 
\textrm{where} \quad I_j =  b_j \sum_{i = 1, i \neq j}^{N} w_{ji} f_2(x_i)$, as defined in 
eq.(\ref{eq:DiffSgmNN_e_signal_elementwise_v2}). 

Using eq.(\ref{eq:DiffSgmNN_e_signal_v2}), we get 
  
\begin{equation} \label{eq:prf1}
{\bf e}^{k+1} - {\bf e}^{k} = 
\left[
\begin{array}{c}
0  \\
0  \\
\vdots  \\
-a_{11} ( x_j^{k+1} - x_j^{k} ) \\
\vdots  \\
0 
\end{array}
\right]
+
\left[
\begin{array}{c}
w_{1j}  \\
w_{2j}  \\
\vdots  \\
0 \\
\vdots  \\
w_{Nj} 
\end{array}
\right]
\big( f_2(x_j^{k+1}) - f_2(x_j^{k}) \big)
\end{equation}

Using the error signal definition of eq.(\ref{eq:DiffSgmNN_e_signal_elementwise_v2}) in 
eq.(\ref{eq:Diff_SgmCIR_elementwise}) gives 

\begin{equation} \label{eq:prf2}
x_j^{k+1} - x_j^{k} = \alpha f_1(e_j^k)  \quad 
\end{equation}

So, the error signal for state $j$ is obtained using 
eq.(\ref{eq:prf1}) and 
(\ref{eq:prf2}) as follows

\begin{eqnarray} 
e_j^{k+1} - e_j^k & = & -a_{jj} ( x_j^{k+1} - x_j^{k} )  \label{eq:prf3} \\
                  & = & -a_{jj} \alpha f_1(e_j^k) \label{eq:prf3_b}
\end{eqnarray}

From eq.(\ref{eq:prf3}) and (\ref{eq:prf3_b}), if $\alpha$ is chosen to satisfy 
$|e_j^k| > |a_{jj} \alpha f_1(e_j^k)|$, then 

\begin{equation} \label{eq:prf4}
|e_j^{k+1}| < |e_j^{k}|, \quad \textrm{for}  |e_i^{k}| \neq 0
\end{equation}

Since sigmoid function $f_1(\cdot)$ is an increasing odd function and $f_1(e_j)=0$ if and only 
if $e_j=0$, then it's seen that there  $\alpha$ can easily be chosen small enough to satisfy 
$|e_j^k| > \alpha a_{jj} |f_1(e_j^k)|$ according to the parameter $a_{jj}$ and slope of sigmoid 
function $f_1(\cdot)$. 

Above, we examined only the state $j$ and its error signal $e_j(k)$. In what follows, we examine 
the evolution of the norm of the complete error vector ${\bf e}^{k+1}$ in eq.(\ref{eq:prf1}). 
From the point of view of the $l_1$ norm of the ${\bf e}^{k+1}$, the worst case is that 
while $|e_j^k|$ decreases, all other elements $|e_i^k|, \quad i \neq j$, increases. So, 
using eq.(\ref{eq:prf1}), (\ref{eq:prf3}) and (\ref{eq:prf4}), we obtain 
that:  If 

\begin{equation} \label{eq:prf7} 
|-a_{jj} ( x_j^{k+1} - x_j^{k} )|  \geq 
	|f_2(x_j^{k+1}) - f_2(x_j^k)| \sum_{i=1, (i \neq j)}^{N}  |w_{ij}| 
\end{equation} 

then 

\begin{equation} \label{eq:prf8}
|| {\bf e}(k+1) ||_1 
\left\{ 
\begin{array}{ll}
 < || {\bf e}(k) ||_1  &  \quad \textrm{if} ||{\bf e}(k)||_1 \neq {\bf 0}  \\
 =  0         &  \quad \textrm{if} ||{\bf e}(k)||_1 = {\bf 0} 
\end{array}
\right.
\end{equation}

The sigmoid function $f_2(\cdot)$ is a Lipschitz continuous function as shown in Appendix A.  So, 

\begin{equation} \label{eq:prf_Lipsc} 
k_2 |x_j^{k+1} - x_j^{k}| \geq  |f_2(x_j^{k+1}) - f_2(x_j^k)| 
\end{equation} 

where $k_2 = 0.5 \sigma$ is $f_2(\cdot)$'s global Lipschitz constant as shown in Appendix A. 

From eq.(\ref{eq:prf7}) and (\ref{eq:prf_Lipsc}), choosing 
$| a_{jj} | > k_2 | \sum_{i=1, (i \neq j)}^{N} |w_{ij}|$ yields eq.(\ref{eq:prf7}), which implies 
eq.(\ref{eq:prf8}). This completes the proof. 

\end{proof}

\emph{Proposition 2: } \\

In asynchronous mode, 
choosing the slope of $f_2(\cdot)$ relatively small as compared to $f_1(\cdot)$ and 
choosing $a_{jj}>0$ and $\alpha$ satisfying (\ref{eq:lemma1_if_alpha}), 
the D-Sgm"SIR"NN in eq.(\ref{eq:Diff_SgmCIR_elementwise}) with a symmetric matrix ${\bf W}$ 
is stable and there exists a finite step number $T_d$ such that 
the $l_1$-norm of the error vector in eq.(\ref{eq:perfIndexLk}) 
goes to zero as its steady state. If $\bar{\theta}_i = \theta_{i}^{tgt1} = 1$ at the 
converged point, then it corresponds to a prototype vector as defined above. 



\begin{proof} \label{prf:2b}

Since it's asynchronous mode, eqs.(\ref{eq:prf1})-(\ref{eq:prf4}) holds where $a_{jj}>0$. So, if $\alpha^k$ 
at time $k$ is chosen to satisfy 
$|e_j^k| > |a_{jj} \alpha^k f_1(e_j^k)|$ as in (\ref{eq:lemma1_if_alpha}), then 

\begin{equation} \label{eq:prf4b}
|e_j(k+1)| < |e_j(k)|, \quad \textrm{for} \quad |e_i(k)| \neq 0
\end{equation}

Note that it's straighforward to choose a sufficiently small $\alpha^k$ to satisfy 
(\ref{eq:lemma1_if_alpha}) according to $a_{jj}$ and the slope $\sigma$ of sigmoid $f_1(\cdot)$. 

Using eq.(\ref{eq:prf1}), (\ref{eq:prf3}) and (\ref{eq:prf4b}), it's seen for $e_j^k \neq 0$ that:  If 

\begin{eqnarray} \label{eq:prf7b} 
\label{eq:prf7b} |-a_{jj} ( x_j^{k+1} - x_j^{k} )|  &  =  &  |- a_{jj} \alpha f_1(e_j^k)| \\
 		&  > &  |f_2(x_j^{k+1}) - f_2(x_j^k)| \sum_{i=1, (i \neq j)}^{N}  |w_{ij}| \label{eq:prf7b_b} 
\end{eqnarray} 

then 

\begin{equation} \label{eq:prf8b}
|| {\bf e}(k+1) ||_1 < || {\bf e}(k) ||_1 
\end{equation}

We observe from eq.(\ref{eq:prf3}), (\ref{eq:prf7b}), (\ref{eq:prf7b_b}) and (\ref{eq:prf8b}) that:

1) If the $x_i^{k}, \quad i=1, \dots, N$, approach to either of the saturation regimes of its 
sigmoid function $f_2(\cdot)$, then 

\begin{eqnarray} \label{eq:prf7c} 
|f_2(x_j^{k+1}) - f_2(x_j^k)| \sum_{i=1, (i \neq j)}^{N}  |w_{ij}| \approx 0, \quad \quad j=1, \dots, N
\end{eqnarray} 

since $|f_2(x_j^{k+1}) - f_2(x_j^k)| \approx 0, \quad \quad i=1, \dots, N$. That makes eq.(\ref{eq:prf7b}) and
(\ref{eq:prf7b_b}) hold.  
Therefore, the norm of the error vector in eq.(\ref{eq:perfIndexLk}) does not go to infinity, 
and is finite for any ${\bf x}$.

2) ${\bf x}(k+1) = {\bf x}(k)$ if and only if ${\bf e}(k) = {\bf 0}$, i.e., 

\begin{eqnarray} \label{eq:prf7d} 
x_j^{k+1} = x_j^{k}  \quad \textrm{if and only if} \quad f_1(e_j^k)=0, \quad \quad j=1, \dots, N
\end{eqnarray} 

3) Examining the eq.(\ref{eq:prf2}), (\ref{eq:prf3}) and (\ref{eq:prf3_b}) 
taking the observations 1 and 2 into account, we 
conclude that any of the $x_j^k,  \quad j=1, \dots, N$, does not go to infinity, and is finite for any $k$.
So, the D-Sgm"SIR"NN in eq.(\ref{eq:Diff_SgmCIR_elementwise}) with a symmetric matrix ${\bf W}$ 
is stable for the assumptions in proposition 2. Because there is a finite number of in-saturation states, 
(i.e. the number of all possible in-saturation state combinations is finite), which is equal to $2^N$, 
there exists a finite step number, say $T_d$, such that ${\bf e}(t) = 0$ for any $t \geq T_d$.

From eq.(\ref{eq:cirD}), if $\bar{\theta}_i = \theta_{i}^{tgt1} = 1$ at the converged point, then 
it corresponds to a prototype vector as defined in previous section, which completes the proof. 

\end{proof}

In what follows, we examine the evolution $\bar{\theta}_i^k$.  
From eq.(\ref{eq:cirD}), by choosing $\theta_j^{tgt}=1$ , let's define the following error signal at time $k$ 

\begin{equation} \label{eq:prop3}
\xi_j^k = -\theta_j^k + \theta_j^{tgt} = -\theta_j^{k} + 1, \quad \quad j=1, \dots, N
\end{equation}

\emph{Lemma 1: } \\

In asynchronous mode, in the D-Sgm"SIR"NN in eq.(\ref{eq:Diff_SgmCIR_elementwise}) 
with a sufficiently small $\alpha^k$ and  
with a symmetric matrix ${\bf W}$, 
the $\xi_k$ is getting closer to $\theta_j^{tgt}=1$ at those  iteration steps $k$ 
where $I_j^k \neq 0$, i.e., $|\xi_j(k+1)| < |\xi_j(k)|$, where 
index $j$ shows the state being updated at iteration $k$. 


\begin{proof} \label{prf:2}

Let $j$ shows the state which is updated at time $k$. The fictitious "SIR" is defined by eq.(\ref{eq:cirD}) 
for nonzero $I_j^k $ as follows 

\begin{equation} \label{eq:prf2_theta}
\bar{\theta}_j^k = \frac{a_{ii} x_j^k}{ I_j^k },  \quad \textrm{where}  \quad 
		I_j^k = b_j + \sum_{i = 1, i \neq j}^{N} w_{ji} f(x_j^k)
\end{equation} 

Let's define the following error signal in $\bar{\theta}_j$, named $\xi_j$, as follows 

\begin{equation} \label{eq:prf2_ksi}
\xi_j^k = -\bar{\theta}_j^k + 1 = \frac{ -a_{ii} x_j^k + I_j^k }{ I_j^k } 
\end{equation}

In asynchronous mode, from eq.(\ref{eq:prf2_theta}), $I_m^{k} = I_m^{k+1}$. 
Using this observation and eq.(\ref{eq:prf2_ksi}) 

\begin{equation} \label{eq:prf2_ksi2}
\xi_j^{k+1} - \xi_j^{k} = \frac{ -a_{ii} (x_j^{k+1} - x_j^{k}) }{ I_j^k } 
\end{equation}

From (\ref{eq:Diff_SgmCIR_elementwise}) and (\ref{eq:prf2_ksi2}), 

\begin{equation} \label{eq:prf2_ksi_diff}
\xi_j^{k+1} - \xi_j^{k} = \frac{ -a_{ii} \alpha f_1(e_j^k) }{ I_j^k } 
\end{equation} 

Provided that $I_j^k \neq 0$, we write from eq.(\ref{eq:DiffSgmNN_e_signal_elementwise_v2}) and 
(\ref{eq:prf2_ksi}), 

\begin{equation} \label{eq:prf2_ksi_e_theta}
e_j^k = I_j^k \xi_j^k
\end{equation} 

Writing eq.(\ref{eq:prf2_ksi_e_theta}) in (\ref{eq:prf2_ksi_diff}) gives 

\begin{equation} \label{eq:prf2_ksi_diff2}
\xi_j^{k+1} - \xi_j^{k} = \frac{ -a_{ii} \alpha f_1( I_j^k \xi_j^k ) }{ I_j^k } 
\end{equation} 

From (\ref{eq:prf2_ksi_diff2}), since sigmoid function $f_1(\cdot)$ is an odd function, and $a_{ii}>0$ and 
$\alpha > 0$, 

\begin{equation} \label{eq:prf2_ksi_diff2delta} 
\xi_j^{k+1} = \xi_j^{k} - \beta sign( \xi_j^{k} ) \quad \textrm{where} \quad 
\beta = | \frac{ -a_{ii} ( \alpha f_1( I_j^k \xi_j^k ) ) }{ I_j^k } | 
\end{equation} 

As seen from eq.(\ref{eq:prf2_ksi_diff2delta}), for a nonzero $\xi_j^{k}$, 
choosing a sufficiently small $\alpha$ satisfying 
$|\xi_j^{k}| > \beta$ assures that 

\begin{equation} \label{eq:prf2_eks_norm} 
|\xi_j^{k+1}| <  |\xi_j^{k}| \quad \textrm{if} \quad I_m^k \neq 0 
\end{equation} 

which completes the proof. 

\end{proof}


Proposition 3: 

In asynchronous mode, 
provided that the D-Sgm"SIR"NN in eq.(\ref{eq:Diff_SgmCIR_elementwise}) 
with a symmetric matrix ${\bf W}$ converges to one of 
the prototype vectors according to proposition 1 and 2,  

\begin{equation} \label{eq:prop3_theta}
\bar{\theta}_i^k = 1,  \quad  j=1, \dots, N
\end{equation}

if and only if $I_i^{k} \neq 0, \quad  j=1, \dots, N$ for the converged prototype vector.
 

\begin{proof} \label{prf:3}

Proposition 1 and 2 shows that the norm of $e_i^k$ decreases and approaches to zero in a finite step number 
and lemma 1 shows the norm of $\xi_i^k$ also decreases if $I_i^k \neq 0$. 
From eq.(\ref{eq:prf2_ksi_e_theta}), $e_i^k = I_i^k \xi_i^k$: 
As the $e_i^k$ approaches to zero, then  
$\xi_i^k$ also approaches to zero, 
given that 
$I_i^k \neq 0$. This is sketched by the 
following equation

\begin{eqnarray} \label{eq:prf2_eks_norm_b} 
\xi_i^{k} = 0,  \quad (i.e., \bar{\theta}_i^k = 1) \quad \quad 
		\textrm{if} \quad e_i^{k}  = 0, \quad \textrm{and} \quad I_i \neq 0 \\
\end{eqnarray} 

On the other hand, if $\xi_i^{k} = 0$ at the converged fixed point, then $e_i^k = 0$ because 
from eq.(\ref{eq:prf2_ksi_e_theta}), $e_i^k = I_i^k \xi_i^k$ 

\begin{eqnarray} \label{eq:prf2_eks_norm_c} 
\quad e_i^{k}  = 0 \quad \quad  \textrm{if} \quad \xi_i^{k} = 0 
\end{eqnarray} 

provided that $I_i \neq 0, \quad  j=1, \dots, N$, which completes the proof. 

\end{proof}

Proposition 4:

The results in proposition 1 and 2 for asynchronous mode hold also for synchronous mode.

\begin{proof} \label{prf:3}

In asynchronous mode, from eq.(\ref{eq:DiffSgmNN_e_signal_v2}) 

\begin{equation} \label{eq:prf3_1}
{\bf e}^{k+1} - {\bf e}^{k} = 
\sum_{i=1}^{N}
\Big(
\left[
\begin{array}{c}
0  \\
0  \\
\vdots  \\
-a_{11} ( x_i^{k+1} - x_i^{k} ) \\
\vdots  \\
0 
\end{array}
\right]
+
\left[
\begin{array}{c}
w_{1i}  \\
w_{2i}  \\
\vdots  \\
0 \\
\vdots  \\
w_{Ni} 
\end{array}
\right]
( f_2(x_i^{k+1}) - f_2(x_i^{k}) )
\Big)
\end{equation}

Using (\ref{eq:DiffSgmNN_e_signal_elementwise_v2}) in eq.(\ref{eq:prf3_1}) and 
writing elementwise gives 

\begin{equation} \label{eq:prf3_2}
e_i^{k+1} = e_i^k - a_{ii} \alpha f_1(e_i^k) + \sum_{j=1, (j \neq i)}^{N}  w_{ij} ( f_2(x_j^{k+1} - f_2(x_j^k) ), 
\quad i=1, \dots, N
\end{equation}

From eq.(\ref{eq:prf3_1}) and (\ref{eq:prf3_2}), we obtain 

\begin{eqnarray} 
\label{eq:prf4_2a} |-a_{ii} ( x_i^{k+1} - x_i^{k} )| &  = & |- a_{ii} \alpha f_1(e_i^k)| \\
			& > & |f_2(x_i^{k+1}) - f_2(x_i^k)| \sum_{j=1, (j \neq i)}^{N}  |w_{ji}| 
\quad i=1, \dots, N
\label{eq:prf4_2a_b} 
\end{eqnarray}

which is equal to (\ref{eq:prf7}) in proposition 1 and (\ref{eq:prf7b}) in proposition 2.  
Continueing the the steps of the analysis in proposition 1 and proposition 2 yield the results 
in proposition 1 and proposition 2 respectively.

\end{proof}

\subsection{Fixed-Step Discrete``SIR''NN Network (FS"SIR"NN) \label{FS}}

In this subsection, establishing an analogy to the traditional fixed step 1-bit 
increase/decrease power control algorithm e.g. \cite{Herdtner00}, \cite{Kim01}, 
we propose the following network by 
replacing the $f_1(\cdot)$ in eq.(\ref{eq:Diff_SgmCIR_elementwise}) by sign function 
as shown in the following 

\begin{equation} \label{eq:Sign''SIR''NN_elementwise} 
x_{j}^{k+1} = x_{j}^{k} + \Delta sign \Big(
-a_{jj} x_j^k + b_j + \sum_{i = 1, i \neq j}^{N} w_{ji} f_2(x_j^k) \Big)
\quad \quad j=1, \dots, N
\end{equation}

where $f_2(\cdot)$ represents the sigmoid function.  
We call the network in eq.(\ref{eq:Sign''SIR''NN_elementwise}) as Fixed-Step ``SIR'' Neural Networks (FS''SIR''NN).

\vspace{0.2cm}  
Corollary 1:
\vspace{0.2cm}

In the FS''SIR''NN in eq.(\ref{eq:Sign''SIR''NN_elementwise}) with a symmetric matrix ${\bf W}$, 
the $l_1$-norm of the error vector in eq.(\ref{eq:perfIndexLk}) converges to 
the interval $[-a_{ii} \alpha, +a_{ii} \alpha]$ while the $x_i^k$ converges to 
the interval $[\alpha, +\alpha]$ 
within a finite step number in asynchoronous mode if 

\begin{equation} \label{eq:lemma1_if}
| a_{jj} | \geq k_2  \sum_{i=1, (i \neq j)}^{N}  |w_{ij}|
\end{equation}

where $k_2= 0.5 \sigma$ is the the global Lipschitz constant
of $f_2(\cdot)$  as shown the in Appendix A.

\begin{proof} \label{prf:corr1}

\vspace{0.2cm}

We're going to obtain the results by writing $f_1(\cdot) = sign(\cdot)$ in the proof of proposition 1 
in section \ref{FDPC} above. 
This would correspond to a sigmoid function $f_1(\cdot)$ whose slope is infinity in proposition 1. 
  
Let $j$ show the state which is updated at time $k$. 
Following the steps in eq.(\ref{eq:DiffSgmNN_e_signal_v2}) and (\ref{eq:prf1}) 
and writing $f_1(\cdot) = sign(\cdot)$ in eq.(\ref{eq:prf2}) gives 

\begin{equation} \label{eq:prop5_a}
x_j^{k+1} - x_j^{k} = \alpha sign(e_j^k)
\end{equation} 

So, the error signal for state $j$ is obtained using 
eq.(\ref{eq:prf1}) and \ref{eq:prop5_a} as follows 

\begin{eqnarray} 
e_j^{k+1} - e_j^k & = & -a_{jj} ( x_j^{k+1} - x_j^{k} )  \label{eq:prop5_b_b} \\
                  & = & -a_{jj} \alpha sign(e_j^k) \label{eq:prop5_b_c}
\end{eqnarray}

From eq.(\ref{eq:prop5_b_b}) and (\ref{eq:prop5_b_c}), 

\begin{equation} \label{eq:prop5_d}
|e_j^{k+1}|
\left\{ 
\begin{array}{ll}
 < |e_j^{k}|  &  \quad \textrm{if} |e_j^{k}| > |\alpha a_{jj}| \\
 < |\alpha a_{jj}|   &  \quad \textrm{otherwise} 
\end{array}
\right.
\end{equation}

Above, we examined only the state $j$ and its error signal $e_j(k)$. In what follows, we examine 
the evolution of the norm of the complete error vector ${\bf e}^{k+1}$ in eq.(\ref{eq:prf1}). 
From the point of view of the $l_1$ norm of the ${\bf e}^{k+1}$, the worst case is that 
while $|e_j^k|$ decreases, all other elements $|e_i^k|, \quad i \neq j$, increases. So, 
using eq.(\ref{eq:prf1}), eq.(\ref{eq:prop5_b_b}-(\ref{eq:prop5_d}), we obtain 
that:  If 

\begin{equation} \label{eq:prop5_e}
|-a_{jj} ( x_j^{k+1} - x_j^{k} )|  \geq 
	|f_2(x_j^{k+1}) - f_2(x_j^k)| \sum_{i=1, (i \neq j)}^{N}  |w_{ij}| 
\end{equation} 

then 

\begin{equation} \label{eq:prop5_f}
|| {\bf e}(k+1) ||_1 
\left\{ 
\begin{array}{ll}
 < || {\bf e}(k) ||_1  &  \quad \textrm{if} ||{\bf e}(k)||_1 > \Delta \sum_{i=1, (i \neq j)}^{N} |a_{ii}|  \\
 < \Delta \sum_{i=1, (i \neq j)}^{N} |a_{ii}|  &  \quad \textrm{otherwise}
\end{array}
\right.
\end{equation}

The sigmoid function $f_2(\cdot)$ is a Lipschitz continuous function as shown in Appendix A, 
$k_2 |x_j^{k+1} - x_j^{k}| \geq |f_2(x_j^{k+1}) - f_2(x_j^k)|$, 
where $k_2 = 0.5 \sigma$ is $f_2(\cdot)$. From eq.(\ref{eq:prop5_e}) and the Lipschitz inequality, 
choosing $| a_{jj} | > k_2 | \sum_{i=1, (i \neq j)}^{N} |w_{ij}|$ satisfies eq.(\ref{eq:prop5_e}), 
which implies eq.(\ref{eq:prop5_f}). This completes the proof. 

\end{proof}

\vspace{0.2cm}
Corollary 2:
\vspace{0.2cm}

In asynchronous mode, 
choosing $f_1(e_i)=sign(e_i)$, $a_{jj}>0$ and $\alpha > 0$, 
the FS"SIR"NN in eq.(\ref{eq:Sign''SIR''NN_elementwise}) 
with a symmetric matrix ${\bf W}$ 
is stable and there exists a finite step number $T_d$ such that 
the $l_1$-norm of the error vector in eq.(\ref{eq:perfIndexLk}) 
converges to 
the interval $[-a_{ii} \alpha, +a_{ii} \alpha]$ while the $x_i^k$ 
converges to the interval $[\alpha, +\alpha]$ 
within a finite step number.

\begin{proof} \label{prf:corr2}

Writing $f_1(\cdot) = sign(\cdot)$ in proposition 2 in section \ref{FDPC} and following the steps and the 
observations therein gives that the FS"SIR"NN in eq.(\ref{eq:Sign''SIR''NN_elementwise}) 
is stable and there exists a finite step number $T_d$ such that 
the $l_1$-norm of the error vector in eq.(\ref{eq:perfIndexLk}) 
converges to 
the interval $[-a_{ii} \alpha, +a_{ii} \alpha]$ while the $x_i^k$ 
converges to the interval $[\alpha, +\alpha]$. 

\end{proof}

\vspace{0.2cm}
Corollary 3:  
\vspace{0.2cm}

The results in corollary 1 and 2 for asynchronous mode hold also for synchronous mode.

\begin{proof} \label{prf:corr3}

Writing $f_1(\cdot) = sign(\cdot)$ in proposition 1 and 2 in section \ref{FDPC} 
and following the the steps and observations of the analysis as in proposition 4 
in section \ref{FDPC} for synchronous mode yields the results 
in corollary 1 and 2 respectively above.

\end{proof}

It's known form literature that the performance of Hopfield network may highly depend on the 
parameter setting of the weight matrix (eg. \cite{Muezzinoglu05}). 
There are various ways for determining the weight matrix of the Hopfield Networks: Gradient-descent
supervised learning (e.g. \cite{Haykin99}),
solving linear inequalities (e.g. \cite{Berg96}, \cite{Harrer91} among others),
Hebb learning rule \cite{Hebb49}, \cite{Muezzinoglu04} etc.
How to design CINR-SgmNN 
is out of the scope of this paper. The methods used for traditional Hopfied NN can also be used for the 
proposed networks D-Sgm``CIR''NN and FS``CIR''NN.



\smallskip
\smallskip

\section{Simulation Results  \label{Section:SimuResults} }

We take the same examples as in \cite{Uykan08a} for comparison reasons and 
for the sake of brevity and easy reproduction of the simulation results.  
In \cite{Uykan08a}, the performances of continuous-time networks, Sgm"SIR"NN and Hopfield networks,
are examined.  In this paper, their discrete-time versions are examined. 
We apply the same Hebb-based (outer-products-based) design procedure (\cite{Hebb49}) in \cite{Uykan08a}, 
which is presented in Appendix B in case of orthogonal prototype vectors.  

In this section, we present two examples, one with 8 neurons and one with 16 neurons. 
The weight matrices are designed by the outer products-based design in Appendix B.

As in \cite{Uykan08a}, traditional Hopfield network is used a reference network.  
The discrete Hopfield Network \cite{Hopfield85} is 

\begin{equation} \label{eq:discreteHopfield}
{\bf x}^{k+1} = sign \Big( {\bf W} {\bf x}^{k}  \Big)
\end{equation} 

where ${\bf W}$ is the weight matrix and ${\bf x}^{k}$ is the state at time $k$, and at most one state is 
updated. 

\vspace{0.2cm}
\emph{Example 1:}
\vspace{0.2cm}

In this example, there are 8 neurons.  The desired prototype vectors are 

\begin{equation} \label{eq:ex1_D}
{\mathbf D} =
\left[
\begin{array}{c c c c c c c c}
1   &   1   & 1  &  1  & -1  &  -1  & -1  & -1  \\
1   &   1   & -1  &  -1  & 1  &  1  & -1  & -1  \\
1   &   -1  &  1  &  -1  & 1  &  -1  & 1  & -1 
\end{array}
\right]
\end{equation}

The weight matrices $\bf{ A }$ and $\bf{ W }$, and the threshold vector $\bf{ b }$
are obtained as follows
by using the outer-products-based design presented in Appendix B and
the slopes of sigmoid functions $f_1(\cdot)$ and $f_2(\cdot)$ are set to $\sigma_1=10$ and
$\sigma_2=2$ respectively, and $\rho=0$, $\alpha = 0.1$ and $\Delta = 0.1$.

\begin{equation} \label{eq:matA_W_b_ex1}
{\mathbf A} = 3 {\mathbf I},  
\quad \quad
{\mathbf W} =
\left[
\begin{array}{c c c c c c c c}
0   &   1   &  1  &  -1  &  1  &  -1 & -1  & -3 \\
1   &   0   & -1  &   1  & -1  &  1  & -3  & -1 \\
1   &   -1  &  0  &   1  & -1  &  -3 & 1  & -1 \\
-1  &   1   & 1   &  0   & -3  &  -1 & -1 &  1 \\
1   &   -1  & -1  &  -3  & 0  &  1   &  1  & -1 \\
-1  &   1   &  -3 &  -1  & 1  &  0   & -1  & 1 \\
-1  &   -3  & 1   &  -1  &  1 &  -1  &  0  & 1 \\
-3  &   -1  & -1  &   1  & -1 &  1  &  1  & 0
\end{array}
\right],  
\quad \quad
{\mathbf b} = {\mathbf 0}
\end{equation}

\begin{figure}[htbp]
  \begin{center}
   \epsfxsize=30.0em    
\leavevmode\epsffile{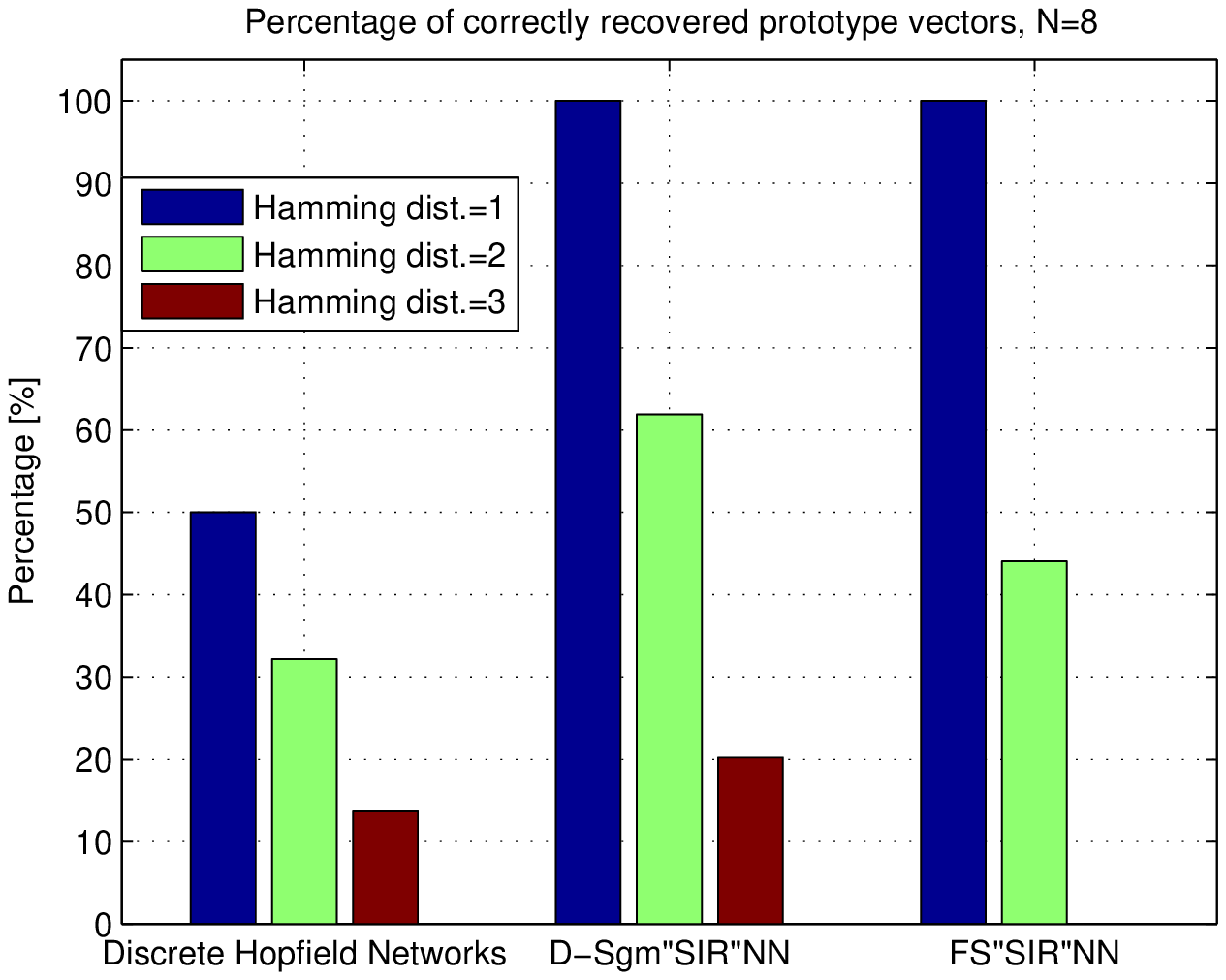}
   \vspace{-1em}        
  \end{center}
 \caption[ The figure 
shows percentage of correctly recovered desired patterns for 
all possible initial conditions in example 1 for the proposed D-Sgm"SIR"NN and Sign"SIR"NN 
as compared to traditional Hopfield network with 8 neurons. ]{  }
\label{fig:CIRHop_ex1_percentage} 
\end{figure}

The Figure \ref{fig:CIRHop_ex1_percentage} shows the percentages of correctly recovered desired patterns for
all possible initial conditions $\mathbf{ x }^k \in (-1,+1)^N$, in the proposed networks D-Sgm''SIR''NN 
and FS-Sgm''SIR''NN 
as compared to traditional discrete Hopfield network. 

Let $m_d$ show the number of prototype vectors and $C(N,K)$, (such that $N \geq K \geq 0$), represent the
combination $N, K$, which is equal to $C(N,K)=\frac{N!}{(N-K)! K!}$, where $!$ shows factorial.
In our simulation, the prototype vectors are from $(-1,1)^N$ as seen above. For initial conditions,
we alter the sign of $K$ states where $K$=0, 1, 2, 3 and 4, which means the initial condition
is within $K$-Hamming distance from the corresponding prototype vector.
So, the total number of different possible combinations for the initial conditions for this example is
24, 84 and 168 for 1, 2 and 3-Hamming distance cases respectively, which
could be calculated by $m_d \times C(8,K)$, where $m_d =3$ and $K=$ 1, 2 and 3.

As seen from Figure \ref{fig:CIRHop_ex1_percentage} the performance of the proposed network D-Sgm"SIR"NN 
is remarkably better than that of traditinal discrete Hopfield NN for 1, 2 and 3 Hamming distance cases. 
The FS''SIR''NN also considerably outperforms the Hopfield for 1 and 2 Hamming distance cases while 
Hopfield NN outperforms FS''SIR''NN at 3 Hamming distance case.

\vspace{0.2cm}
\emph{Example 2:}
\vspace{0.2cm}

The desired prototype vectors are 

\begin{equation} \label{eq:ex1_D}
{\mathbf D} =
\left[
\begin{array}{c c c c c c c c c c c c c c c c}
1 & 1 & 1 & 1 & 1 & 1 & 1 & 1 & -1 & -1 & -1 & -1 & -1 & -1 & -1 & -1 \\
1 & 1 & 1 & 1 & -1 & -1 & -1 & -1 & 1 & 1 & 1 & 1 & -1 & -1 & -1 & -1 \\
1 & 1 & -1 & -1 & 1 & 1 & -1 & -1 & 1 & 1 & -1 & -1 & 1 & 1 & -1 & -1 \\
1 & -1 & 1 & -1 & 1 & -1 & 1 & -1 & 1 & -1 & 1 & -1 & 1 & -1 & 1 & -1
\end{array}
\right]
\end{equation}

The weight matrices ${\bf A}$ and ${\bf W}$ and 
threshold vector ${\bf b}$ is obtained as follows 
by using the outer products based design explained above.

\begin{eqnarray} \label{eq:matA_W_b_ex2}
{\mathbf A} & = & 4 {\mathbf I}, \nonumber \\
{\mathbf W} & = &
\left[
\begin{array}{c c c c c c c c c c c c c c c c}
 0  &  2  &  2  &  0  &  2  &  0  &  0  & -2  &  2  &  0  &  0  & -2  &  0  & -2  & -2  & -4 \\
 2  &  0  &  0  &  2  &  0  &  2  & -2  &  0  &  0  &  2  & -2  &  0  & -2  &  0  & -4  & -2 \\
 2  &  0  &  0  &  2  &  0  & -2  &  2  &  0  &  0  & -2  &  2  &  0  & -2  & -4  &  0  & -2 \\
 0  &  2  &  2  &  0  & -2  &  0  &  0  &  2  & -2  &  0  &  0  &  2  & -4  & -2  & -2  &  0 \\
 2  &  0  &  0  & -2  &  0  &  2  &  2  &  0  &  0  & -2  & -2  & -4  &  2  &  0  &  0  & -2 \\
 0  &  2  & -2  &  0  &  2  &  0  &  0  &  2  & -2  &  0  & -4  & -2  &  0  &  2  & -2  &  0 \\
 0  & -2  &  2  &  0  &  2  &  0  &  0  &  2  & -2  & -4  &  0  & -2  &  0  & -2  &  2  &  0 \\
-2  &  0  &  0  &  2  &  0  &  2  &  2  &  0  & -4  & -2  & -2  &  0  & -2  &  0  &  0  &  2 \\
 2  &  0  &  0  & -2  &  0  & -2  & -2  & -4  &  0  &  2  &  2  &  0  &  2  &  0  &  0  & -2 \\
 0  &  2  & -2  &  0  & -2  &  0  & -4  & -2  &  2  &  0  &  0  &  2  &  0  &  2  & -2  &  0 \\
 0  & -2  &  2  &  0  & -2  & -4  &  0  & -2  &  2  &  0  &  0  &  2  &  0  & -2  &  2  &  0 \\
-2  &  0  &  0  &  2  & -4  & -2  & -2  &  0  &  0  &  2  &  2  &  0  & -2  &  0  &  0  &  2 \\
 0  & -2  & -2  & -4  &  2  &  0  &  0  & -2  &  2  &  0  &  0  & -2  &  0  &  2  &  2  &  0 \\
-2  &  0  & -4  & -2  &  0  &  2  & -2  &  0  &  0  &  2  & -2  &  0  &  2  &  0  &  0  &  2 \\
-2  & -4  &  0  & -2  &  0  & -2  &  2  &  0  &  0  & -2  &  2  &  0  &  2  &  0  &  0  &  2 \\
-4  & -2  & -2  &  0  & -2  &  0  &  0  &  2  & -2  &  0  &  0  &  2  &  0  &  2  &  2  &  0
\end{array}
\right],   \nonumber \\
{\mathbf b} & = & {\mathbf 0}
\end{eqnarray}

\begin{figure}[htbp]
  \begin{center}
   \epsfxsize=30.0em    
\leavevmode\epsffile{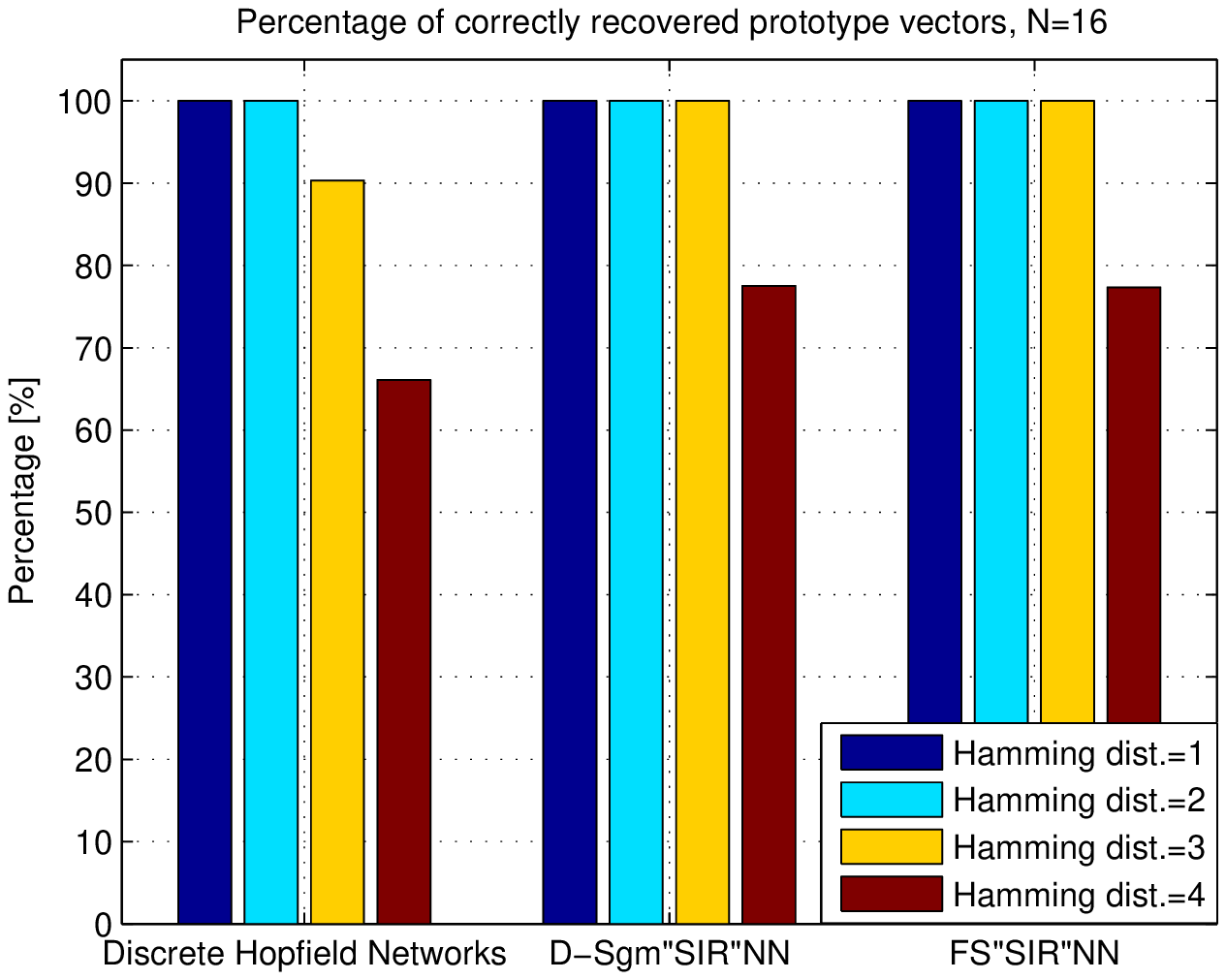}
   \vspace{-1em}        
  \end{center}
 \caption[ The figure 
shows percentage of correctly recovered desired patterns for 
all possible initial conditions in example 2 for the proposed D-Sgm"SIR"NN and Sign"SIR"NN 
as compared to traditional Hopfield network with 16 neurons. ]{  }
\label{fig:CIRHop_ex2_percentage} 
\end{figure}

The Figure \ref{fig:CIRHop_ex2_percentage} shows percentage of correctly recovered desired patterns for 
all possible initial conditions $\mathbf{ x }^k \in (-1,+1)^16$, in the proposed D-Sgm"SIR"NN  and 
FS''SIR''NN
as compared to discrete Hopfield network.  

The total number of different possible combinations for the initial conditions for this example is
64, 480 and 2240 and 7280 for 1, 2, 3 and 4-Hamming distance cases respectively, which
could be calculated by $m_d \times C(16,K)$, where $m_d =4$ and $K=$ 1, 2, 3 and 4.

As seen from Figure \ref{fig:CIRHop_ex2_percentage} the performance of the proposed networks D-Sgm"SIR"NN and 
FS"SIR"NN is the same as that of discrete Hopfield Network for 1-Hamming and 2-Hamming distance cases 
($\%100$ for all networks).  However, 
the D-Sgm"SIR"NN and FS"SIR"NN gives better performance than the discrete Hopfield network does for 3 and 4 
Hamming distance cases.

\section{Concluding Remarks  \label{Section:CONCLUSIONS}}

This paper is continuation of the work in \cite{Uykan08a} where 
we present and analyse a Sigmoid-based 
"Signal-to-Interference Ratio, (SIR)" balancing dynamic 
network in continuous time. 
In this second part, we present the corresponding 
network in discrete time:  
We show that 
in the proposed discrete-time network, called D-Sgm"SIR"NN, the defined error vector approaches 
to zero in a finite step in both synchronous and asynchronous work modes.  
Our investigations show that 
i) Establishing an analogy to the distributed (sigmoid) power control algorithm in 
\cite{UykanPhD01} and \cite{Uykan04}, 
if the defined fictitious "SIR" is equal to 1 at the converged eqiulibrium point, then it is one of the 
prototype vectors. 
ii) The D-Sgm"SIR"NN exhibits similar features as discrete-time Hopfield NN does. 
iii) Establishing an analogy to the traditional 1-bit fixed-step power control algorithm, the corresponding 
"1-bit" network, called Sign"SIR"NN network, is also presented. 
Computer simulations show the effectiveness of the
proposed networks as compared to traditional discrete Hopfield Network.

\smallskip
\smallskip

\section*{Appendix A}

In what follows, we will show the sigmoid function 
($f_2(a) = 1 - \frac{2}{1 + exp(-\sigma a)}, \quad \sigma>0$) has the global Lipschitz constant
$k = 0.5 \sigma$.

Since $f ( \cdot )$ is a differentiable function, we can apply the mean value theorem 

\begin{eqnarray} \label{eq:nebiliim}
 f(a) - f(b) = (a-b) f^{'}( \mu a + (1- \mu) (b-a))  \nonumber \\
with \quad \mu \in [0,1]  \nonumber
\end{eqnarray}

The derivative of $f(\cdot)$ is $f^{'}(a) = \frac{\sigma}{e^{\sigma a} {(1+e^{\sigma a})^2} }$ 
whose maximum is at the point $a = 0$, i.e., $| f^{'}(a) | \leq 0.5 \sigma$.  So we obtain the 
following inequality 

\begin{equation}\label{eq:atkafadan}
| f(a) - f(b) | \leq k |a-b|  
\end{equation}

where $k = 0.5 \sigma$ is the global Lipschitz constant of the sigmoid function.  

\vspace{0.2cm}

\section*{Appendix B}


\emph{ Outer products based network design: }
\vspace{0.2cm}

Let's assume that $L$ desired orthogonal prototype vectors, $\{ \mathbf{ d }_s \}_{s=1}^{L}$, 
are chosen form  $(-1, +1)^N$.


Step 1: Calculate the sum of outer products of the prototype vectors (Hebb Rule, \cite{Hebb49})

\begin{eqnarray} \label{eq:HebbQd}
\mathbf{ Q } = \sum_{s=1}^{L} \mathbf{ d }_s  \mathbf{ d }_s^T
\end{eqnarray}

Step 2: Determine the diagonal matrix $\bf{A}$ and $\bf{W}$ as follows:

\begin{equation} \label{eq:AfromHebb}
a_{ij} = 
\left\{
\begin{array}{ll}
q_{ii} + \rho & \textrm{if} \quad i = j, \\
0 & \textrm{if} \quad i \neq j
\end{array}
\right.  \quad \quad i,j=1, \dots, N
\end{equation}

where $\rho$ is a real number and   

\begin{equation} \label{eq:WfromHebb}
w_{ij} =
\left\{
\begin{array}{ll}
0 & \textrm{if} \quad i = j, \\
q_{ij} & \textrm{if} \quad i \neq j
\end{array}
\right.   \quad \quad i,j=1, \dots, N
\end{equation}

where $q_{ij}$ shows the entries of matrix $\mathbf{ Q }$, $N$ is the dimension of the vector 
$\mathbf{ x }$ and $L$ is the number of the prototype vectors ($N > L > 0$). In eq.(\ref{eq:AfromHebb}), 
$q_{ii} = L$ from (\ref{eq:HebbQd}) since $\{ \mathbf{ d }_s \}$ is from $(-1, +1)^N$ and 
$\rho$ is a real number.  However, from the analysis in section \ref{FDPC} and \ref{FS},  
it can be seen that the 
proposed networks D-Sgm"SIR"NN and FS"SIR"NN contain the prototype vectors 
as their equilibrium points for a relatively large interval of $\rho$. 

Another choice of $\rho$ in (\ref{eq:AfromHebb}) is $\rho = N-2L$ which yields $a_{ii} = N-L$.  
In what follows we show that this choice also assures that $\{ \mathbf{ d }_j  \}_{j=1}^L$ 
are the equilibrium points of the networks. 


From (\ref{eq:HebbQd})-(\ref{eq:WfromHebb})

\begin{equation} \label{eq:HebbDPCA}
[ - \mathbf{ A } + \mathbf{ W } ] = -(N-L) \mathbf{ I } + 
    \sum_{s=1}^{L} \mathbf{ d }_s  \mathbf{ d }_s^T - L  \mathbf{ I } 
\end{equation}

where $\mathbf{ I }$ represents the identity matrix. 

Since $\mathbf{ d }_s \in (-1,+1)^N$, then  $|| \mathbf{ d }_s ||_2^2 = N$.  
Using (\ref{eq:HebbDPCA}) and the orthogonality 
properties of the set $\{ \mathbf{ d }_s  \}_{s=1}^L$ gives 

\begin{equation} \label{eq:HebbDPCA_a}
[ - \mathbf{ A } + \mathbf{ W } ] \mathbf{ d }_s  = -(N-L) \mathbf{ d }_s  + (N-L) 
	\mathbf{ d }_s  = \mathbf{ 0 }
\end{equation} 

So, the prototype vectors $\{ \mathbf{ d }_j \}_{j=1}^L$ correspond to equilibrium points.

\section*{Acknowledgments}

This work was supported in part by Academy of Finland and Research Foundation (Tukis\"{a}\"{a}ti\"{o}) of Helsinki
University of Technology, Finland.

\nocite{*}
\bibliographystyle{IEEE}

%

\vspace{3cm}


\begin{thebibliography}{100}

\bibitem{Hopfield85}
J.J. Hopfield and D.W Tank,
\newblock Neural computation of decisions in optimization problems
\newblock {\em Biological Cybernetics}, vol. :141-146, 1985.

\bibitem{Matsuda98}
S. Matsuda,
\newblock ``Optimal'' Hopfield network for combinatorial optimization with linear cost function,
\newblock {\em IEEE Trans. Neural Networks}, vol. 9: 1319-1330, Nov. 1998.

\bibitem{Smith98}
K. Smith, M. Palaniswami, and M. Krishnamoorthy,
\newblock Neural techniques for combinatorial optimization with applications,
\newblock {\em IEEE Trans. Neural Networks}, vol. 9: 1301-1318, Nov. 1998.

\bibitem{Tan05}
K.C. Tan, T. Huajin and S.S. Ge,
\newblock On parameter settings of Hopfield networks applied to traveling salesman problems,
\newblock {\em Circuits and Systems I}, vol. 52, nr. 5: 994-1002, May 2005.

\bibitem{Paik92}
J.K. Paik and A.K. Katsaggelos,
\newblock Image restoration using a modified Hopfield network,
\newblock {\em IEEE Trans. Image Processing}, vol. 1, nr. 1:49-63,  Jan. 1992.

\bibitem{Rappaport96} T.S. Rappaport,
\newblock{\em Wireless Communications: Principles and Practice},
\newblock Prentice-Hall, New York, 1996.

\bibitem{Farrel90}
J.A. Farrel and A.N. Michel,
\newblock A synthesis procedure for Hofield's continuous-time associative memory,
\newblock {\em IEEE Trans. Circuits Systems}, vol. 37: 877 - 884, 1990.

\bibitem{Lendaris99}
G.G. Lendaris, K. Mathia and R. Saeks,
\newblock Linear Hopfield networks and constrained optimization
\newblock {\em IEEE Trans. Systems, Man, and Cybernetics, Part B}, vol. 29, nr. 1: 114 - 118 Feb. 1999.

\bibitem{Haykin99}
S. Haykin,
\newblock{\em Neural Networks},
\newblock Macmillan, 1999.


\bibitem{UykanPhD01}
Z. Uykan,
\newblock {\em Clustering-based algorithms for Radial Basis Function and Sigmoid
Perceptron Networks}.
\newblock PhD thesis, Control Eng. Lab., Helsinki University of Technology, 2001.

\bibitem{Uykan04} Z. Uykan and H.N. Koivo,
\newblock ``A sigmoid basis nonlinear power control algorithm for mobile
radio systems'',
\newblock IEEE Trans. Vehic. Tech., 2003

\bibitem{Uykan08a} Z. Uykan,
\newblock ``From Sigmoid Power Control Algorithm to Hopfield-like Neural Networks:
``SIR'' (``Signal''-to-``Interference''-Ratio)-
Balancing Sigmoid-Based Networks- Part I:  Continuous Time'',
\newblock {\em submitted to IEEE Trans. Neural Networks, 2008}.

\bibitem{Zurada92}
J.M. Zurada,
\newblock{\em Introduction to Artificial Neural Systems},
\newblock West Publishing Company, 1992.

\bibitem{Muezzinoglu04}
M.K. Muezzinoglu and C. Guzelis,
\newblock A Boolean Hebb rule for binary associative memory design,
\newblock {\em IEEE Trans. Neural Networks}, vol. 15, nr. 1:195 - 202, Jan. 2004.

\bibitem{Muezzinoglu05}
M.K. Muezzinoglu, C. Guzelis and J.M. Zurada,
\newblock An energy function-based design method for discrete hopfield associative memory
with attractive fixed points
\newblock {\em IEEE Trans. Neural Networks}, vol. 16, nr. 2:370-378, March 2005 .



1995.







\bibitem{Huajin04}
T. Huajin, K.C. Tan and Y. Zhang,
\newblock A columnar competitive model for solving combinatorial optimization problems,
\newblock {\em IEEE Trans. Neural Networks}, vol. 15, nr. 6: 1568 - 1574, Nov. 2004.

\bibitem{Zander92a} J. Zander,
\newblock ``Performance of optimum transmitter power control
in cellular radio systems,''
\newblock {\it IEEE Trans. Veh. Technol.}, vol.
VT-41, pp. 57-62, 1992.

\bibitem{Zander92b} J. Zander,
\newblock ``Distributed cochannel interference control in
cellular radio systems'',
\newblock {\it IEEE Trans. Veh. Technol.}, vol.
VT-41, pp. 305-311, 1992.



\bibitem{Hebb49} D. O. Hebb ,
\newblock{\em The Organization of Behaviour },
\newblock John Wiley and Sons, New York, 1949.


\bibitem{Berg96} J. Van den Berg,
\newblock ``The most general framework of continuous Hopfield neural networks'',
\newblock Proc. Int. Workshop on of Neural Networks for Identification, Control,
Robotics, and Signal/Image Processing, pp. 92 - 100, 21-23 Aug. 1996.


\bibitem{Harrer91} H. Harrer, J.A. Nossek and F. Zou,
\newblock ``A learning algorithm for time-discrete cellular neural networks'',
\newblock Proc. IEEE Int. Joint Conf. on Neural Networks, vol.1, pp. 717 - 722, 1991.

\bibitem{Herdtner00}
J.D. Herdtner and E.K.P. Chong,
\newblock Analysis of a class of distributed asynchronous power control algorithms for cellular wireless systems, 
\newblock {\em IEEE Journal on Selected Areas in Comm.}, 
18(3): 436 - 446, March 2000.

\bibitem{Kim01}
Dongwoo Kim;
\newblock On the convergence of fixed-step power control algorithms with binary feedback for mobile communication systems, 
\newblock {\em IEEE Transactions on Communications},
49(2): 249 - 252, Feb 2001.


\end{thebibliography}


\newpage

\listoffigures

\newpage

\newpage

\end{document}